\documentclass[twocolumn,prl]{revtex4}

\usepackage{graphicx}

\begin{document}
\title{Single Stranded DNA Translocation Through A Nanopore:

 A Master Equation Approach
}

\author{O. Flomenbom and J. Klafter}
\address{\it School of Chemistry, Raymond \& Beverly Sackler Faculty Of Exact Sciences,\\  Tel-Aviv University, \it
Tel-Aviv 69978, Israel}

\date{\today}

\begin{abstract}
We study voltage driven translocation of a single stranded (ss)
DNA through a membrane channel. Our model, based on a master
equation (ME) approach, investigates the probability density
function ({\it pdf}) of the translocation times, and shows that it
can be either double or mono-peaked, depending on the system
parameters. We show that the most probable translocation time is
proportional to the polymer length, and inversely proportional to
the first or second power of the voltage, depending on the initial
conditions. The model recovers experimental observations on
hetro-polymers when using their properties inside the pore, such
as stiffness and polymer-pore interaction.
\end{abstract}
\maketitle
\begin{center}
{\bf {1. INTRODUCTION}}\end{center}

Translocation of biopolymers through a pore embedded in a membrane
is a fundamental step in a variety of biological processes. Among
the examples are the translocation of some species of m-RNA
through the nucleus membrane which is the first stage of gene
expression in eucaryotic cells [1], and the attack of cells by
viruses that occurs by injecting the genetic information through a
hole in the cell membrane [2]. The translocation importance in
bio-systems, and the possibility for developing fast sequencing
methods, have been the motivation for recent experiments, in which
a voltage-driven ssDNA is translocated through
$\alpha$-$hemolysin$ channel of a known structure [3]. In these
experiments one measures the time it takes a single ssDNA molecule
to pass through a membrane channel [4-7]. Since ssDNA is
negatively charged (each monomer has an effective charge of $zq$,
where $z$ ($0<z<1$ ) is controlled by the solution pH and
strength), when applying a voltage the polymer is subject to a
driving force while passing through the membrane from the negative
(cis) side to the positive (trans) side. Because the presence of
the ssDNA in the transmembrane pore part ({\it {TPP}}) blocks the
cross-{\it {TPP}} current, one can deduce the translocation times
{\it pdf} from the current blockade duration times~[4-7]. It has
been found that the shape of the translocation times {\it pdf} is
controlled not only by the voltage applied to the system, the
temperature and the polymer length but also by the nature of the
homo-polymer used: poly-$dA$ ($A$-adenine); poly-$dC$
($C$-cytosine); poly-$T_{nu}$ ($T_{nu}$-thymine), and the sequence
of hetro-polymers [4-7].

The translocation process can be roughly divided into two stages.
The first stage is the arrival of the polymer in the vicinity of
the pore and the second stage is the translocation itself. Several
models have been suggested for describing the translocation stage.
In [8] an equation for the free energy of the translocation,
obtained from the partition functions of the polymer parts outside
the {\it {TPP}}, was derived and used to calculate the mean first
passage time (MFPT). Other investigators used similar ideas with
improved free energy terms by taking into consideration effects
such as the membrane width [9-10], or assumed that only the part
of the ssDNA inside the {\it {TPP}} affects the dynamics of the
translocation rather than the polymer parts outside the {\it
{TPP}} [11-12].

In this work we present a new theoretical approach that allows to
consider both the polymer parts outside the {\it {TPP}} and within
the {\it {TPP}}. Using the ME we are able to map the three
dimensional translocation onto a discrete space one-dimensional
process. Based on the ME we compute the {\it pdf} of the first
passage times (FPT) of the translocation, $F(t)$, and the MFPT, as
a function of the system parameters. We relate our theoretical
results to recent experimental observations and by analyzing them
using our model, we come up with physical understanding of these
observations.
\begin{center}
{\bf 2. THE MODEL}\end{center}

An $n(=N+d-1)$-state ME is used to describe the translocation of
an $N$-monomer long ssDNA subject to an external voltage $V$, and
temperature $T$, through a {\it {TPP}} of a length that
corresponds to $d(=12)$ monomers. A state is defined by the number
of monomers on each side of the membrane and within the {\it TPP}.
A change in the state of the system (only nearest states
transitions are allowed) is assumed to be controlled mainly by the
behavior of the polymer within the {\it TPP} in the presence of
the applied voltage. In addition, it is assumed to be influenced
by entropic and interaction factors of the polymer outside and
within the {\it TPP}. Absorbing ends are used as boundary
conditions, which are the natural choice for this system because
the polymer can exit the {\it TPP} on both sides. The state $j=n$
represents the arrival of the first monomer into the {\it TPP}
from the cis side, and the state $j=0$ represents the departure of
the far end monomer from the trans side of the {\it TPP}. Let
$P_j(t)$ be the {\it pdf} of state $j$ that contains $m_j$
nucleotides occupying the {\it TPP} according to:
 \begin{eqnarray}
  m_j=
 \left\{%
 \begin{array}{cccc}
    j, & j\leq d & \nonumber & ~~~~~~~(1.1) \\
    d, & N>j>d & given~N\geq d, & ~~~~~~~(1.2) \\
    n+1-j, & j\geq N & \nonumber & ~~~~~~~(1.3) \\
 \end{array}%
 \right.
 \end{eqnarray}
and a similar set of equations for a short polymer, $N\leq d$,
which is obtained by exchanging $N$ and $d$ in Eqs.(1.1) through
(1.3). The governing equations of motion are:
\renewcommand{\theequation}{2}
\begin{eqnarray}
 {\partial} P_j(t)/{\partial}
 t=a_{j+1,j}P_{j+1}(t)+a_{j-1,j}P_{j-1}(t)~~~~~~~~~~~~~~~~~~{\nonumber}\\
-(a_{j,j+1}+a_{j,j-1})P_{j}(t);~~j=2,\cdot \cdot \cdot,
n-1,{\nonumber}\\
\partial P_{y}(t)/\partial t=a_{y+1,y}P_{y+1}(t)\delta_{y,1}+a_{y-1,y}P_{y-1}(t)\delta_{y,n}{\nonumber}~~~~~~~\\
-(a_{y,y-1}+a_{y,y+1})P_{y}(t);~~y=1,n.~~~~~~ (2){\nonumber}
\end{eqnarray}
Equations (2) can be written in matrix representation $\partial
\overrightarrow{P}/\partial t = \mathbf{A} \overrightarrow{P}$.
The propagation matrix $\mathbf{A}$, is a tridiagonal matrix that
contains information about the transitions between states in term
of rate constants. We assume that the rate constants can be
decoupled into two terms:
\renewcommand{\theequation}{3}
\begin{equation}\
a_{j,j+/-1}=k_{j}(T)p_{j,j+/-1}(V,T).
 \end{equation}
The first term provides the rate to perform a step, while the
second term gives the probability to jump from state $j$ in a
given direction, and obeys: $p_{j,j+1}+p_{j,j-1}=1$. To obtain
$k_{j}$, we first consider the bulk relaxation time of a polymer
[13] $\tau_b\propto \beta\xi_b b^2 N^\mu$ , where
$\beta^{-1}=k_BT$,~{\it b} is a monomer length,~$\xi_b$ is the
Stokes bulk friction constant per segment, $\xi_b=6\pi b
\eta$~(where $\eta$ is the solvent viscosity), $N$ is the number
of monomers in the polymer and the dimensions dependant $\mu$
represents the effect of the microscopic repulsion on the average
polymer size. In three dimensions $\mu =3,9/5,3/2$ for rod-like,
self-avoiding and Gaussian (Zimm model) chains, respectively. To
compute the relaxation time inside the {\it TPP}, the confined
volume of the {\it TPP} must be taken into account. For a rod-like
polymer the restricted volume dictates a one-dimensional
translocation, whereas for a flexible polymer the limitations are
less restricted. We implement this by taking $\mu$ as a measure of
the polymer stiffness inside the {\it TPP} that obeys:~$0\leq
\mu\leq 1.5$. The expression for the relaxation rate of state j is
therefore,
\renewcommand{\theequation}{4}\begin{equation}
 k_j=1/(\beta
\xi_{p}b^2m_{j}^\mu)\equiv R/m_{j}^\mu.
\end{equation}
From Eq.(4) it is clear that as $\mu$ becomes smaller the rate to
preform a step becomes larger, namely $k_j$ for a rod-like polymer
increases. This feature appears, at first sight, to be in
contradiction to the relaxation time behavior of a bulk polymer,
where a rod-like polymer has a larger relaxation time than that of
a flexible polymer. This contradiction is resolved by taking into
account the different dimensional demand for a rod-like polymer
relative to a flexible polymer inside the {\it TPP}. Because $\mu$
is a measure of the polymer stiffness inside the {\it TPP}, it is
controlled by the interaction between the monomers occupying the
{\it TPP}, e.g base stacking and hydrogen bonds, and therefore is
affected by the monomer type and the sequence of the ssDNA.

The friction constant per segment inside the {\it TPP}, $\xi_{p}$,
represents the interaction between the ssDNA and the {\it
TPP}.~The physical picture is that during translocation there are
few or no water molecules between the polymer and the {\it TPP}.
Consequently the water molecules inside the {\it {TPP}} can hardly
be viewed as the conventional solvent and the Stokes friction
constant is replaced by $\xi_p$ representing the ssDNA-{\it TPP}
interaction.

To calculate $p_{j,j-1}$, the second term on the right hand side
of Eq.(3), we assume a quasi-equilibrium process and use the
detailed balance condition for the ratio of the rate constants
between neighboring states: $a_{j,j-1}/a_{j-1,j}=e^{-\beta\Delta
E_j}$, where $\Delta E_j=E_{j-1}-E_{j}$. We then use the
approximation $a_{j,j-1}/a_{j-1,j}\approx
p_{j,j-1}/(1-p_{j,j-1})$, and deduce the jump probabilities by
computing $\Delta E_j$. To compute $E_j$ the contributions from
three different sources are considered: electrostatic, $E^p_j$,
entropic, $E^s_j$, and an averaged interaction energy between the
ssDNA and the pore, $E^i_j$.

For the calculations of the electrostatic energy difference
between adjacent states, $\Delta E_j^p$, we assume a linear drop
of the voltage along the {\it TPP} and write for $m_j$ monomers
occupying the {\it TPP} penetrating from the cis side of the
membrane for a length of $i_jb$:
\renewcommand{\theequation}{5}
\begin{equation}\label{5}
  E^p_j=zq(V/d)\sum_{n=i_j}^{m_j+i_j-1}n=zq(V/2d)m_j(m_j+2i_j-1).
\end{equation}
The effective charge per monomer in the {\it TPP} is taken to be
the same as of the bulk.  For states that contain monomers at the
trans side of the membrane, $zqV$ should be added for $\Delta
E_j^p$. This contribution represents the additional effective
charge that passed through the potential $V$. Accordingly, the
expression for $\Delta E_j^p$ is (see Appendix A):
\renewcommand{\theequation}{6}
\begin{equation}\label{2}
  \Delta E_j^p=zqV(m_j+\alpha_j)/d.
\end{equation}
Here $\alpha_j$ takes the values: $\alpha_j=\{-1;0;1\}$ for cases
described by equations $\{(1.1); (1.2); (1.3)\}$ respectively
($\alpha_j=1$ and $\alpha_j=-1$ correspond to the entrance and
exit states of the translocation, whereas $\alpha_j=0$ corresponds
to the intermediate states of the translocation). For a short
polymer, $\alpha_j$ has the same values as for a large polymer.

The contribution to $\Delta E_j$ from $\Delta E_j^s$ is composed
of two terms. One term is the entropic cost needed to store $m_j$
monomers inside the {\it TPP}, and the second term originates from
the reduced number of configurations of a Gaussian polymer near an
impermeable wall. Combining these terms leads to (see Appendix B):
\renewcommand{\theequation}{7}
\begin{equation}\label{7}
   \Delta E^s_j=\gamma_j/\beta,
\end{equation}
where $\gamma_j=\{-1+g_{j};g_{j};1+g_{j}\}$, for cases described
by equations $\{(1.1); (1.2); (1.3)\}$, respectively. $g_{j}$ is
given in Appendix B in terms of $N_{j,cis}$ and $N_{j,trans}$,
which are the number of monomers on the cis and trans sides
correspondingly. For a short polymer $\gamma_j$ behaves similarly
but for intermediate states $g_{j}$=$0$.

For computing $\Delta E_j^i$ we focus on the average interaction
between the ssDNA and pore (not only its transmembrane part). Due
to the asymmetry of the pore between the cis and the tran sides of
the membrane [3], the ssDNA interacts with the pore on the cis
side of the membrane and within the {\it TPP} but not on the trans
side of the membrane. Assuming an attractive interaction,
$E_j^i=-k_BT(N-N_{j,trans})$ (more properly,
$E_j^i=-\varepsilon(N-N_{j,trans})$, and we set $\epsilon=k_BT$ in
the relevant temperature regime) we obtain:
\renewcommand{\theequation}{8}
\begin{equation}\label{8}
   \Delta E^i_j=\zeta_j/\beta,
\end{equation}
where $\zeta_j=\{1;1;0\}$, for the cases described by equations
$\{(1.1) ; (1.2); (1.3)\}$, respectively, and for a short polymer
$\zeta_j=\{1;0;0\}$.

Using the above relations, and defining
$\delta_j=\gamma_j+\zeta_j$ we obtain
\renewcommand{\theequation}{9}
\begin{equation}
  p_{j,j-1}=(1+e^{\beta \Delta E_j^p+\delta_j})^{-1}.
\end{equation}
For the system to be voltage driven $\beta|\Delta E_j^p|>\delta_j$
must be fulfilled, which translate into the condition: $V/V_C>1$,
where a characteristic voltage is introduced: $V_C^{-1}\equiv
(1+1/d)\beta z|q|$. This inequality ensures that there is a bias
towards the trans side of the membrane. Otherwise the polymer is
more likely to exit from the same side it entered than to
transverse the membrane. Under experimental conditions [6]
$V_C=46mV$, when using $z\approx1/2$.

In {\it Figure 1} we show the different contributions to $\Delta
E_j$: $\beta \Delta E_j^p$ (for $\beta z|q|V$=$1$) and $\delta_j$.
$\beta\Delta E_j^p$ decreases for the entrance states of the
translocation, increases at the exit states of the translocation,
and is a negative constant for intermediate states. Clearly
$\beta\Delta E_j^p\leq 0$ reflects the field directionality. On
the other hand, $\delta_j$ opposes the translocation for the
entrance and intermediate states. For the entrance states
$\delta_j>0$ due to both entropic terms, but approaches zero (from
below) for the exit states of the translocation, due to the
cancellation of $\Delta E_j^i$ against the entropic gain of
storing less monomers within the {\it TPP}. At intermediate states
$\delta_j\approx 1$, where its shape near the crossover between
the different situations is controlled by $g_{j}$.
\begin{center}
{\bf {3. RESULTS AND DISCUSSION}}\end{center}
\begin{center}
{\bf 3.i. The FPT {\it pdf}}\end{center}

In this subsection we compute the FPT {\it pdf}, $F(t)$, and
examine its behavior as a function of the system parameters.
$F(t)$ is defined by:
\renewcommand{\theequation}{10}
\begin{equation}\label{10}
 F(t)=\partial(1-S(t))/\partial t,
\end{equation}
where the survival probability, namely, the probability to have at
least one monomer in the {\it TPP}, $S(t)$, is the sum over all
states {\it pdf}
\renewcommand{\theequation}{11}
\begin{equation}\label{11}
  S(t)=\overrightarrow{U}\mathbf{C}e^{\mathbf{D}t}\mathbf{C}^{-1}\overrightarrow{P}_0.
\end{equation}
Here $\overrightarrow{U}$ is the summation row vector of $n$
dimensions, $\overrightarrow{P}_0$ is the initial condition column
vector,~$(\overrightarrow{P}_0)_j=\delta_{x,j}$, where $x$ is the
initial state, and the definite negative real part eigenvalue
matrix, $\mathbf{D}$, is obtained through the transformation:
$\mathbf{D}=\mathbf{C}^{-1}\mathbf{A}\mathbf{C}$.
\begin{figure}[t]
\includegraphics[width=0.85\linewidth,angle=0]{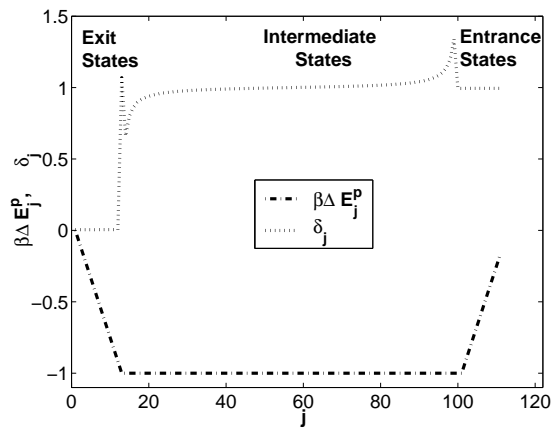}
\caption{The components of the $\Delta E_j$, $\beta\Delta E_j^p$
($\beta z|q|V=1$) and $\delta_j$, are shown for $N=100$.}
\end{figure}

Substituting Eq.(11) into Eq.(10), $F(t)$ is obtained. {\it Figure
2} shows the double-peaked $F(t)$ behavior as a function of
$V/V_C$ and $N$ (inset), for starting at $x=N+d/4$. The left peak
represents the non-translocated events. Its amplitude decreases as
$V/V_C$ increases but remains unchanged (along with the position)
with the increase in $N$, because only the 'head' of the polymer
is involved in these events. As $x$ decreases, namely when the
translocation initial state shifts towards the trans side, the
non-translocation peak and the translocation peak merge and for
$x=n/2$, namely, for an initial condition for which
$N_{j,cis}$=$N_{j,trans}$, $F(t)$ has one peak (data not shown)
independent of $V/V_{C}$.
\begin{figure}[t]
\includegraphics[width=0.85\linewidth,angle=0]{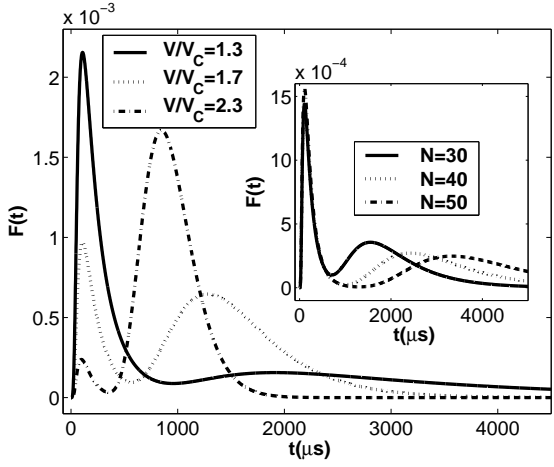}
\caption{$F(t)$, for several values of $V/V_C$, with: $N$=$30$,
$x$=$N+d/4$, $z\approx 1/2$, $T$=$2^oC$, $V_C$=$46mV$,
$\mu$=$1.14$ and $R$=$10^6$Hz. The left peak represents the
non-translocated events, whereas the right peak represents the
translocation. Inset: $F(t)$ as a function of $N$, with
$V/V_C=1.5$, and the other parameters as above. The
non-translocation peak remains unchanged when increasing the
polymer length, while the translocation peak shifts to the right
with $N$.}
\end{figure}

For a convenient comparison to experimental results, we calculate
and examine the behavior of the most probable average
translocation velocity, $v=xb/t_m$, where $t_m$ is the time that
maximizes the translocation peak.

$Figure~3$ shows the $v(N)$ behavior, for $V/V_C=2.60$.~$v(N)$ is
a monotoniaclly decreasing function of $N$ for $N\leq d$, but is
independent of $N$ for $N\geq d$. This agrees with the
experimental results [6]. The origin for $v(N)$ behavior, stems
from Eqs.(1) and (4), according to which the minimal $k_j$ is a
decreasing monotonic function of $N$ for $N\leq d$, but is
independent of $N$ for $N\geq d$.

The velocity, $v(V)$, depends on $x$, $v(V)=v_x(V)$, and changes
from a linear to quadratic function of the voltage when changing
$x$, as shown in the inset of $figure~3$, for $N=30$.

Starting at $x=N+1$, a linear scaling is obtained:
$v_{N+1}(V)=b_1(V-b_2/b_1)$.~The coefficient $b_2/b_1$ can be
identified as an effective characteristic voltage:
$\widetilde{V_C}^{-1}\equiv(1+1/d)\beta \widetilde{z}|q|=b_1/b_2$.
From the last equality~$\widetilde{z}$ can be extracted.

Starting at $x=N/2$, i.e. when initial state is close to the exit
states, a square dependence is obtained:
$v_{N/2}(V)=c_1(V-c_2)^2+c_3$,~with $c_1=o(10^{-5})$, $c_2=40mV$
and $c_3=o(10^{-2})$. These coefficients are similar to the
measured values [6].

We note that both linear and square scaling behaviors have been
observed experimentally [4,6]. A possible explanation for the
different functional behavior of $v(V)$ might originate from
different data analysis that can be interpreted as having a
different initial condition.

To get numerical values for $\xi_p$ and $\mu$, we use the
experimental data in refs. [6-7], and obtain:~$\mu(C)=1$,
$\mu(A)=1.14$, $\mu(T_{nu})=1.28$, and $\xi_p(A)\approx
10^{-4}meVs/nm^2 $, $\xi_p(C)=\xi_p(T_{nu})=\xi_p(A)/3$. From
these values we find the limit in which the relaxation time of the
polymer parts outside the {\it TPP} can be neglected. We estimate
the maximal bulk number of monomers, $N_{max}$, for which the bulk
relaxation time is much shorter ($5\%$) than of the {\it TPP}
relaxation time. For a poly-$dA$ bulk Zimm chain we get
$N_{max}\approx 271$,~ by taking for the viscosity the value for
water at $2^oC$, $\eta\approx 1.7\cdot 10^{-3}Ns/m^2$. Using this
value, $\xi_b$ can be calculated from the Stokes relation to
be:~$\xi_b\approx 10^{-7}meVs/nm^2$, which is three order of
magnitude smaller than $\xi_p$.
\begin{figure}[t]
\includegraphics[width=0.85\linewidth,angle=0]{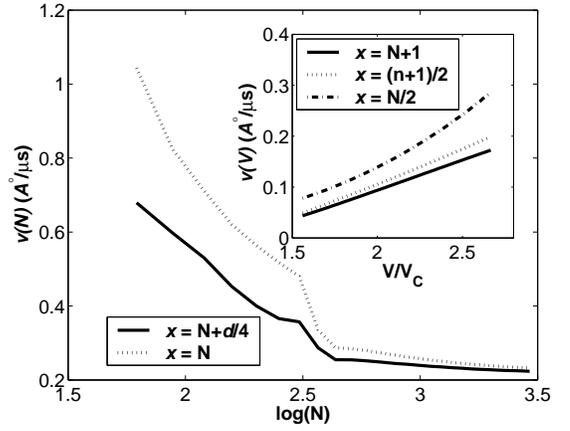}
\caption{The velocity as a function of the polymer length, for the
same parameters as in {\it figure 2}, various initial conditions
(linear-log scales) and $V/V_C=2.6$. $v(N)$ tends towards a length
independent, but display a sharp decrease for short polymers. Note
for the step feature of $v(N)$ when $N$ becomes larger than the
{\it TPP} length, $d$. Inset: $v(V)$ , for $N=30$, showing linear
and quadratic scaling depending on the initial condition, $x$.}
\end{figure}
\\
\begin{center}

{\bf 3.ii. The MFPT}\end{center}

Additional information about the translocation can be obtained by
computing the MFPT, $\overline{\tau}$. To compute an analytical
expression for $\overline{\tau}$, we consider a large polymer,
$N>d$, and replace $p_{j,j-1}$ and $k_j$ by state independent
terms: $p_+=[1+e^{(-V/V_C+1)}]^{-1}$, valid for $x\approx N$, and
$k=1/(\beta \xi_pb^2d^\mu )$. This leads to $a_+=p_+k$ and
$a_-=(1-p_+)k$, which defines a one dimensional state invariant
random walk. The MFPT is obtained by inverting $\mathbf{A}$ [14]:
$\overline{\tau}=\int _0^\infty tF(t)dt=
-\overrightarrow{U}\mathbf{A}^{-1}\overrightarrow{P}_0$. The
calculations of the elements of the state independent
$\mathbf{A}^{-1}$ yields (see Appendix C):
\renewcommand{\theequation}{12}
\begin{equation}\label{12}
  (-\mathbf{A}^{-1})_{s,x}=\frac{\Delta(p^s)\Delta(p^{n+1-x})}{\Delta(p)\Delta(p^{n+1})}\frac{p_+^{x-s}}{k};~~~~~~~s<x,
\end{equation}
where $\Delta (p^m)=p_+^m-p_-^m$ and $(-\mathbf{A}^{-1})_{s,x}$
for $s\geq x$ is obtained when exchanging $s$ with $x$ and $p_+$
with $p_-$ in Eq.(12). Summing the $x$ column elements of
$(-\mathbf{A}^{-1})$ we obtain $\overline{\tau}$ (see Appendix C):
\renewcommand{\theequation}{13}
\begin{equation}
\overline{\tau} = \frac{\Delta (p^{n+1-x})p_+^xx-\Delta
(p^{x})p_-^{n+1-x}(n+1-x)}{k\Delta (p) \Delta (p^{n+1})},
\end{equation}
which in the limit of a weak bias, $V/V_C\gtrsim 1$, can be
rewritten as (see Appendix C)
\renewcommand{\theequation}{14}
\begin{equation}\label{14}
  \overline{\tau}\approx
  \frac{2x\xi_pb^2d^\mu}{z|q|(1+1/d)}\frac{1}{V-V_C}.
\end{equation}
Although $\overline{\tau}$ and $t_m$ are different characteristics
of $F(t)$ and differ significantly when slow translocation events
dominate, Eq.(14) captures the linear scaling with $N$ and $1/V$.
The quadratic scaling of $t_m$ with $1/V$ is obtained when using
Eq.(9) rather than its state invariant version, for starting at,
or near, an initial state for which $\delta_{j=x}\leq0$.
\\
\begin{center}
{\bf 3.iii. The sequence effect}\end{center}

Under the assumptions that are presented below, we now construct
$\xi_p$ and $\mu$ for every ssDNA sequence and thus examine the
sequence effect on $t_m$.~For a given ssDNA sequence occupying the
{\it TPP} in the $j$ state, we write an expression for the average
friction of that state,~$\xi_{p,j}$, assuming additive
contributions of the monomers inside the {\it TPP}:
\renewcommand{\theequation}{15}
\begin{equation}\label{15}
\xi_{p,j}=(1/m_j)\sum_{s=1}^{m_j}\xi_{p}(nu_{s}).
\end{equation}
\begin{figure}[b]
\includegraphics[width=0.85\linewidth,angle=0]{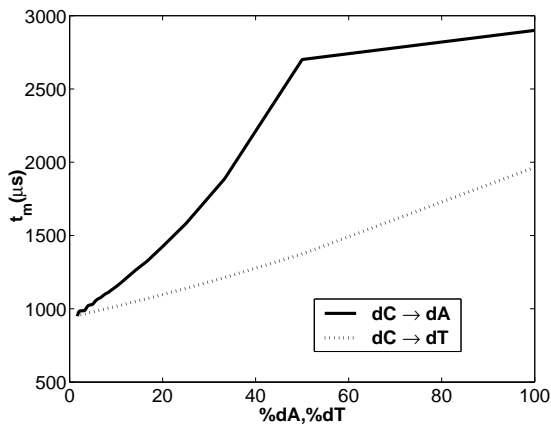}
\caption{$t_m$ as a function of the percentage of equally spaced
monomers substitutions $C\rightarrow A$ and $C\rightarrow T_{nu}$,
for $N=60$, $x=N$, $z\approx 1/2$, $V/V_C=1.63$ and $T=292^oK$.
The curves emphasize the effects of the rigidity and the friction
on the translocation dynamics. See text for discussion.}
\end{figure}
Here $nu_s$ stands for the nucleotide $s$ occupying the {\it TPP}.
To construct a compatible state dependent stiffness parameter,
$\mu_j$, we first argue that only nearest monomers can interact
inside the {\it TPP} and thus contributes to the rigidity of the
polymer, which in turn increases $k_j$. We then examine the
chemical structure of the nucleotides and look for 'hydrogen-like'
bonds between adjacent bases. The term 'hydrogen-like' bonds is
used because the actual distance between the atoms that create the
interaction may be larger than of a typical hydrogen bond. The
pairs $AA$ and $CC$ can interact but not the pairs $T_{nu}T_{nu}$
and $CA$. For the pair $CT_{nu}$ the interaction is orientation
dependent; namely, for the $l$ pairs sequence,
poly-$d(CT_{nu})_l$, the interaction is within each of the pairs
but not between the pairs. Accordingly we have:
$\mu(nu~nu)=\mu(nu)$, $\mu(CA)=\mu(AC)=\mu(T_{nu})$,
$\mu(CT_{nu})=\mu(C)$, $\mu(T_{nu}C)=\mu(T_{nu})$, which allow the
calculation of $\mu_j$ following the definition:
\renewcommand{\theequation}{16}
\begin{equation}\label{16}
  \mu_j=\frac{1}{m_j-1}\sum_{s=1}^{m_j-1}\mu(nu_s~nu_{s+1)},
\end{equation}
with $(\mu_j)_{m_j=1}=0$.~{\it Figure 4} shows $t_m$ as a function
of equally spaced substitutions $C\rightarrow T_{nu}$ and
$C\rightarrow A$. The linear scaling of $t_m(C \rightarrow
T_{nu})$ is due to the linear scaling of $\mu_j(C\rightarrow
T_{nu})$. The saturating behavior of $t_m(C \rightarrow A)$ is a
combination of two opposing factors: the linear scaling of
$\xi_{p,j}(C\rightarrow A)$ and the nonmonotonic behavior of
$\mu_j({C\rightarrow A})$.~Our model also predicts that for
sufficiently large $l$
\renewcommand{\theequation}{17}
\begin{equation}\label{17}
  t_m[(CA)_l]>t_m(C_lA_l).
\end{equation}
This feature is explained by noticing that
$2t_m[(CA)_l]>t_m(C_{2l})+t_m(A_{2l})$, see {\it figure 4}, which
follows from the expression for $\mu_j$, Eq.(16), and then using
$2t_m(C_nA_n)\approx t_m(C_{2l})+t_m(A_{2l})$, which follows from
the linear scaling of $t_m$ with $N$, for $N\geq d$ in addition to
Eq.(15). The above findings regarding the behavior of $t_m$ for
hetro-ssDNA, fit the experimental results [7].
\\
\begin{center}
{\bf 4. CONCLUSIONS}\end{center}

In the presented model, the translocation of ssDNA through
$\alpha$-$hemolysin$ channel is controlled, in addition to the
voltage, by the interaction between the polymer and the pore
($\Delta E_j^i $ and $\xi_p$), and between nearest monomers inside
the {\it TPP} ($\mu$), as well as by an entropic factors originate
from polymer segments outside and within the {\it TPP}. Based on
the model, we showed that $F(t)$ can be mono or double peaked
depending on $x$ and $V/V_C$. We calculated the MFPT to be:
$\overline{\tau}\sim N/(V-V_C)$, for $N>d$ and $V/V_C\gtrsim 1$,
and $t_m\sim N/v_x(V)$ for $N\geq d$, where $v_x(V)$ changes from
a linear to a quadratic function of $V$ with $x$. In addition we
estimated that $\xi_p\approx 10^3\xi_b$, and by constructing
$\xi_p$ and $\mu$ for hetro-ssDNA explained experimental results
regarding the various behaviors of $t_m$ for hetro-ssDNA.

An extended version of this model that describes translocation
through a fluctuating channel structure, can be used to describe
unbiased translocation, which displays long escape times [16].
Translocation of other polymers through proteins channels can be
described using the same framework by changing $\mu$, $\xi_p$ and
$\Delta E_j$.

{\it Acknowledgements:} we acknowledge fruitful discussions with
Amit Meller and with Ralf Metzler and the support of the US-Israel
Binational Science Foundation and the Tel Aviv University
Nanotechnology Center.
\\
\begin{center}
{\bf APPENDIX A}\end{center}

We wish to calculate the electrostatic energy difference between
sates, $\Delta E_j^p$. There are three cases during translocation,
which are described by Eqs.(1.1)-(1.3). For any polymer length,
Eq.(1.1)(exit states) describes a case for which $N_{j,cis}$=$0$
and Eq.(1.3) (entrance states) describes a case for which
$N_{j,trans}$=$0$. For the case described by Eq.(1.2)
(intermediate states), there are monomers on both sides of the
membrane for a large polymer, or no monomers on both sides of the
membrane for a short polymer. Starting from Eq.(5) we have for the
entrance states:
\renewcommand{\theequation}{A1}
\begin{equation}\label{A1}
\Delta E_j^p=\frac{zqV(m_j+1)}{d}.
\end{equation}
For the exit states $\Delta E_j^p$ is composed of two
contributions. One contribution stems from the passage of a
monomer with an effective charge of $zq$ through the potential
$V$:
\renewcommand{\theequation}{A2}
\begin{equation}\label{A2}
   \Delta E_{j,1}^p=zqV.
\end{equation}
The second contribution calculated from Eq.(5) is:
\renewcommand{\theequation}{A3}
\begin{equation}\label{A3}
\Delta E_{j,2}^p=\frac{-i_jzqV}{d}.
\end{equation}
Combining the two contributions we find
\renewcommand{\theequation}{A4}
\begin{equation}\label{A4}
\Delta E_{j}^p=\frac{(d-i_j)zqV}{d}=\frac{(m_j-1)zqV}{d},
\end{equation}
when using $m_j=d+1-i_j$.

For the intermediate states and a large polymer, we have only the
contribution given by Eq.(A2) (the number of monomers within the
{\it TPP} is constant), which can be written as:
\renewcommand{\theequation}{A5}
\begin{equation}\label{A5}
\Delta E_{j}^p=\frac{m_jzqV}{d},
\end{equation}
when using $m_j=d$ which holds for the intermediate states. For a
short polymer, we have to consider only the contribution given by
Eq.(5), which leads again to Eq.(A5). Eq.(6) is obtained from
adding the above contributions.
\\
\begin{center}
{\bf APPENDIX B}\end{center}

For calculating $\Delta E_j^s$, we start by writing an expression
for the entropic energy that consists of two terms:
\renewcommand{\theequation}{B1}
\begin{equation}\label{B1}
    E_j^s=E_{j,1}^s+{E}_{j,2}^s.
\end{equation}
${E}_{j,1}^s$ represents the entropy cost of storing $m_j$
monomers within the {\it TPP} and is a linear function of $m_j$
[10]. $E_{j,2}^s$ originates from the reduced number of
configurations of a Gaussian polymer near an impermeable wall, and
can be approximated by [8]:
\renewcommand{\theequation}{B2}
\begin{equation}\label{B2}
   E_{j,2}^s=
   \left\{%
\begin{array}{ll}
    \frac{1}{2}k_BTln(N_{j,trans}), & j\leq d; \\
    \frac{1}{2}k_BTln(N_{j,trans}N_{j,cis}), & N>j>d; \\
    \frac{1}{2}k_BTln(N_{j,cis}), & j\geq N. \\
\end{array}%
\right.
\end{equation}
Separating the translocation into three regimes, described by
Eqs.(1.1)-(1.3), we find that for the entrance states $\Delta
E_{j,1}^s$ is given by:
\renewcommand{\theequation}{B3}
\begin{equation}\label{B3}
\Delta E_{j,1}^s\propto k_BT,
\end{equation}
while for the exit states
\renewcommand{\theequation}{B4}
\begin{equation}\label{B4}
\Delta E_{j,1}^s\propto -k_BT,
\end{equation}
with a proportional constant of $o(1)$.

For the intermediate states $\Delta E_{j,1}^s=0$ because the same
number of monomers occupy the {\it TPP} between adjacent states.
$\Delta E_{j,1}^s$ for a short polymer has the same values as for
a large polymer, when adjusting the conditions for the three cases
(exchanging $N$ and $d$ in Eqs.(1)).

Computing $\Delta E_{j,2}^s$ from Eq.(B2) results in
$\frac{1}{2}k_BTg_{j}$, where
\renewcommand{\theequation}{B5}
\begin{equation}\label{B5}
   g_{j}=
   \left\{%
\begin{array}{ll}
    ln(1+1/(N-j)), & j\leq d; \\
    ln(1+1/(N-j))(1-1/(j-d)), & N>j>d; \\
    ln(1-1/(j-d)), & j\geq N. \\
\end{array}%
\right.
\end{equation}
For a short polymer, $g_j$ is similar to Eq.(B5) for the entrance
and exit states, but $g_j=0$ for the intermediate states because
$N_{j,cis}$=$N_{j,trans}$=$0$ for these states. Note that
$|g_{j}|<1$ for all $j$. Special care is needed when computing
$g_{j}$ for states that belong to the crossover between the three
situations. For these states a combination of Eqs.(B2) was used.
From the above contributions we obtain Eq.(7).
\\
\begin{center}
{\bf APPENDIX C}\end{center}

To compute $\overline{\tau}$ which is given by
\renewcommand{\theequation}{C1}
\begin{equation}\label{C1}
\overline{\tau}=\sum_{s=1}^{x-1}(-\mathbf{A}^{-1})_{s,x}+\sum_{s=x}^n(-\mathbf{A}^{-1})_{s,x},
 \end{equation}
we have to calculate the elements of the general inverse Toeplitz
matrix $(\mathbf{A}^{-1})_{s,x}$ [17]:
\renewcommand{\theequation}{C2}
\begin{equation}\label{C2}
  (-\mathbf{A}^{-1})_{s,x}=\frac{\Delta(\lambda^s)\Delta(\lambda^{n+1-x})}{\Delta(\lambda)\Delta(\lambda^{n+1})}\frac{p_+^{x-s}}{k};~~~~~~~s<x,
\end{equation}
where $\lambda_{+/-}=[1+/-\sqrt{(1-4p_+p_-)}]/2$. Substituting the
expression for $p_+$ and $p_-$ into the expressions for
$\lambda_{+/-}$, we obtain $\lambda_{+/-}=p_{+/-}$, which when
used in Eq.(C2) results in Eq.(12). The summation of each of the
terms in Eq.(C1) is straight forward. The first term yields
\renewcommand{\theequation}{C3}
\begin{equation}\label{C3}
I=\alpha\sum_{s=1}^{x-1}(1-y^s)=\alpha (x-\frac{1-y^x}{1-y})
 \end{equation}
where $y=p_-/p_+$ and
\renewcommand{\theequation}{C4}
\begin{equation}\label{C4}
   \alpha=\frac{p_+^x\Delta
(p^{n+1-x)}}{\Delta (p) \Delta (p^{n+1})k}.
\end{equation}
The second term in Eq.(C1) is:
\renewcommand{\theequation}{C5}
\begin{eqnarray}\label{C5}
II=\widetilde{\alpha}\sum_{s=x}^{n}\Delta (p^{n+1-s}) p_-^{s-x}=\widetilde{\alpha}p_-^{n+1-x}\cdot~~~~~~~~~~~~~~~~~~~{\nonumber}\\
~~~~~~~~~~\cdot[y^{x-n-1}\frac{1-y^{n+1-x}}{1-y}-(n+1-x)],~~~
 \end{eqnarray}
where
\renewcommand{\theequation}{C6}
\begin{equation}\label{C6}
    \widetilde{\alpha}=\frac{\Delta (p^{x)}}{\Delta (p) \Delta
(p^{n+1})k}.
\end{equation}
Combining {\it I} and {\it II} and rearranging terms results in
Eq.(13).

Rewriting Eq.(13) as
\renewcommand{\theequation}{C7}
\begin{equation}\label{C7}
    \overline{\tau}=\frac{x}{k \Delta
    (p)}-\frac{(n+1)(1-y^{-x})}{k\Delta (p)(1-y^{-n})},
\end{equation}
we find that for $V/V_C\gtrsim 1$, the second term in Eq.(C7)
vanishes as $y^{n-x}$. Keeping the first term in the expression
for $\overline{\tau}$, Eq.(C7), and expanding to first order in
$V/V_C$ the explicit form of $\Delta (p)$
\renewcommand{\theequation}{C8}
\begin{equation}\label{C8}
\frac{1}{\Delta (p)}=\frac{1+e^{-V/V_C+1}}{1-e^{-V/V_C+1}}\approx
\frac{2}{V/V_C-1}
\end{equation}
Eq.(14) is obtained.

\end{document}